\documentclass[a4paper]{article}

\usepackage{INTERSPEECH2022}
\usepackage{xcolor}

\DeclareMathOperator*{\E}{\mathbb{E}}

\title{Minimum Latency Training of Sequence Transducers\\for Streaming End-to-End Speech Recognition}
\name{Yusuke Shinohara$^1$, Shinji Watanabe$^2$}
\address{
  $^1$Yahoo Japan Corporation, Japan, 
  $^2$Carnegie Mellon University, USA}
\email{yusshino@yahoo-corp.jp}

\begin{document}

\maketitle
\begin{abstract}
Sequence transducers, such as the RNN-T and the Conformer-T, are one of the most promising models of end-to-end speech recognition, especially in streaming scenarios where both latency and accuracy are important.
Although various methods, such as alignment-restricted training and FastEmit, have been studied to reduce the latency, latency reduction is often accompanied with a significant degradation in accuracy. We argue that this suboptimal performance might be caused because none of the prior methods \textit{explicitly} model and reduce the latency. 
In this paper, we propose a new training method to explicitly model and reduce the latency of sequence transducer models. First, we define the expected latency at each diagonal line on the lattice, and show that its gradient can be computed efficiently within the forward-backward algorithm. Then we augment the transducer loss with this expected latency, so that an optimal trade-off between latency and accuracy is achieved.
Experimental results on the WSJ dataset show that the proposed minimum latency training reduces the latency of causal Conformer-T from 220 ms to 27 ms within a WER degradation of 0.7\%, and outperforms conventional alignment-restricted training (110 ms) and FastEmit (67 ms) methods.
\end{abstract}
\noindent\textbf{Index Terms}: speech recognition, end-to-end, streaming, sequence transducer, latency

\section{Introduction} \label{sec:introduction}

End-to-end speech recognition is becoming more and more popular in both academia and industry due to the simplicity of the approach as well as the competitive performance relative to conventional hybrid models  \cite{Graves06-CTC,Graves12-STW,Graves13-SRD,Chorowski15-ABM,Watanabe17-HCA,Zeyer18-ITO,Sainath20-SOD,Li20-DRT,Li21-RAE}. Sequence transducer models, such as the recurrent neural network transducer (RNN-T) \cite{Graves12-STW,Graves13-SRD}, the Transformer transducer \cite{Zhang20-TTA,Yeh19-TTE}, and the Conformer transducer (Conformer-T) \cite{Gulati20-CCA,Chen21-DRT}, are one of the most promising end-to-end models, especially in streaming scenarios because of its inherently streaming nature. In streaming speech recognition, latency is one of the primary performance metrics along with recognition accuracy, because a lower latency leads to a quick response of voice-enabled applications, and improves the user experience \cite{Shangguan21-DUP}. However, streaming transducer models tend to delay label emission so as to see more future context to predict labels more accurately, which leads to a large latency and a deteriorated user experience. Hence, much effort has been devoted to reduce the latency of sequence transducer models.

Alignment restricted training \cite{Mahadeokar21-ARS,Sainath20-EWT} is one of the most popular methods for reducing latency of sequence transducers. During the training of sequence transducers, it restricts the alignment between input and output sequences within a certain range from a reference alignment, i.e.~those alignments that are too late (or too early) from the reference are pruned away. As a result, late emission of tokens is prohibited and latency reduction is achieved.
FastEmit \cite{Yu21-FLL} is another well-known method of latency reduction for transducer models. It boosts the gradient of the label emission probability by a specified factor during training, so that the model is encouraged to emit labels earlier. 
It has been empirically shown that those methods considerably reduce the latency of sequence transducer models.

However, these conventional methods often have a significant degradation in recognition accuracy as a side effect of latency reduction \cite{Mahadeokar21-ARS,Sainath20-EWT,Kim21-RSA}\footnote{Some studies have reported no degradation (or even improvement) in accuracy \cite{Yu21-FLL,Li21-BFE}, but we could not reproduce it on our setup.}. We argue that one of the reasons for the suboptimal latency-accuracy trade-off is that none of the prior methods directly model and reduce the latency of transducer models. For instance, alignment restricted training prohibits those alignments that are too late, but it does not compute the gradient of the latency to explicitly reduce the latency in coordination with the transducer loss.

In this paper, we propose a new training method to explicitly model and reduce the expected latency of sequence transducer models. First, we define the expected latency on each diagonal line on the lattice, and show that the expected latency and its gradient can be calculated efficiently within the forward-backward algorithm. Then we propose to augment the transducer loss with this expected latency, so that the transducer loss and the expected latency are reduced in coordination with each other, and an optimal latency-accuracy trade-off is achieved.
Experimental results on the Wall Street Journal (WSJ) dataset show that the proposed minimum latency training achieves a better latency-accuracy trade-off compared with alignment restricted training and FastEmit.

\section{Related Work}

Along with alignment-restricted training and FastEmit, self alignment \cite{Kim21-RSA} is another method of latency reduction for sequence transducers; self alignment is similar to our method in its spirit to adaptively modify gradients depending on locations, rather than uniformly boosting gradients like FastEmit, but different from ours in that it does not explicitly model latency and its gradient. It is worthwhile to note that block-processing encoders \cite{Shi21-EEM,Chen21-DRT} and joint endpointing \cite{Chang19-JEA,Li20-TFA} have been also studied to achieve fast and accurate transducer models.
There is another line of research on streaming end-to-end speech recognition using attention-based encoder-decoder (AED) models. Although AED models \cite{Chorowski15-ABM,Watanabe17-HCA,Zeyer18-ITO} are inherently non-streaming because of the attention mechanism, it can be extended to streaming models \cite{Chiu18-MCA,Moritz19-TAF,Tsunoo21-STA,Inaguma20-MLT}. Among the streaming AED models and its training methods, minimum latency training (MinLT) \cite{Inaguma20-MLT} is similar to our method in that it explicitly models the latency and uses its gradient. However, its formulation is completely different from ours because of the difference in architectures (AED vs.~sequence transducer).

\section{Sequence Transducers \cite{Graves12-STW}}

The sequence transducer is a neural network model to map an input sequence, $X \triangleq \{\bm{x}_1, \dots, \bm{x}_T\}$, to an output sequence, $Y \triangleq \{y_1, \dots, y_U\}$, where $\bm{x}_t \in \mathbb{R}^{d_\texttt{i}}$ is the $d_{\texttt{i}}$-dimensional input vector (e.g.~filter-bank vector) at time $t$, $y_u \in \mathcal{V}$ is the output token (e.g.~character) at position $u$. $T$ is the input sequence length, $U$ is the output sequence length, and $\mathcal{V}$ represents the vocabulary (i.e.~set of tokens). Let $\bar{\mathcal{V}}$ be the extended vocabulary defined as $\mathcal{V} \cup \{\phi\}$, where $\phi$ is a special token to represent a blank. Let alignment $a \in \bar{\mathcal{V}}^{*}$ \footnote{The star symbol represents the Kleene closure, i.e.~$\bar{\mathcal{V}}^{*}$ is a set of all sequences consisting of zero or more tokens in $\bar{\mathcal{V}}$.} represents an alignment between input and output sequences, for instance $a = \{\phi, y_1, \phi, \phi, y_2, y_3, \phi\}$. The sequence transducer defines a conditional distribution of the alignment $P(a | X)$, which is summed over all possible alignments to define
\begin{equation}
P(Y | X) \triangleq \sum_{a \in \mathcal{B}^{-1}(Y)} P(a | X),
\end{equation}
where $\mathcal{B}: \bar{\mathcal{V}}^{*} \rightarrow \mathcal{V}^{*}$ is a function to remove the blank tokens from the alignment.
The sequence transducer consists of an encoder to generate $d_\texttt{e}$-dimensional encoder representation $\bm{f}_t \in \mathbb{R}^{d_\texttt{e}}$ from $\bm{x}_{1:t}$, a predictor network to generate $d_{\texttt{p}}$-dimensional predictor representation $\bm{g}_u \in \mathbb{R}^{d_\texttt{p}}$ from $y_{0:u}$, and a joint network to define a conditional distribution $P(k | t, u)$ of token $k \in \bar{\mathcal{V}}$ given $\bm{f}_t$ and $\bm{g}_u$. The conditional distribution of an alignment $P(a | X)$ is calculated by the product of probabilities $P(k | t, u)$ accumulated over the path defined by the alignment $a$.
The loss function of the sequence transducer model is defined as the negative log probability as
\begin{equation}
\mathcal{L}_{\texttt{trans}} \triangleq -\log P(Y | X).
\label{eq:transducer_loss}
\end{equation}
Its gradients with respect to the label emission probability $P_{y_{u + 1}} \triangleq P(y_{u + 1} | t, u)$ and the blank emission probability $P_{\phi} \triangleq P(\phi | t, u)$ for $\forall t \in \{1, \dots, T\}, \forall u \in \{0, \dots, U\}$ can be calculated as,
\begin{align}
\frac{\partial \mathcal{L}_{\texttt{trans}}}{\partial P_{y_{u+1}}} = - \frac{\alpha(t, u) \beta(t, u + 1)}{P(Y | X)}, \label{eq:trans_loss_grad_y} \\
\frac{\partial \mathcal{L}_{\texttt{trans}}}{\partial P_{\phi}} = - \frac{\alpha(t, u) \beta(t + 1, u)}{P(Y | X)}, \label{eq:trans_loss_grad_b}
\end{align}
where the forward variable $\alpha(t, u)$ and the backward variable $\beta(t, u)$ can be calculated efficiently by the forward and backward recursions as
\footnotesize
\begin{align}
    \alpha(t, u) &= \alpha(t - 1, u) P(\phi | t - 1, u) + \alpha(t, u - 1) P (y_u | t, u - 1), \label{eq:forward_recursion}\\
    \beta(t, u) &= \beta(t + 1, u) P(\phi | t, u) + \beta(t, u + 1) P(y_{u+1} | t, u). \label{eq:backward_recursion}
\end{align}
\normalsize
Note that the initial conditions are set as $\alpha(1, 0) = 1$ and $\beta(T + 1, U) = 1$. This procedure is also known as the \textit{forward-backward algorithm}.

\section{Conventional Methods}

\subsection{Alignment restricted training \cite{Mahadeokar21-ARS}}

Given a reference alignment, typically generated by forced-alignment using an HMM-hybrid model, alignment restricted training prohibits those alignments that are too late (or too early) compared with the reference, with the expectation that the model learns to emit labels not too late than the reference timing. Specifically, the forward recursion in (\ref{eq:forward_recursion}) is modified as
\begin{align}
    \alpha(t, u) &= \alpha(t - 1, u) P(\phi | t - 1, u) m_\phi(t, u) \nonumber \\
                 &~~~~~ + \alpha(t, u - 1) P (y_u | t, u - 1) m_y(t, u),
\end{align}
where $m_\cdot(t, u) \in \{0, 1\}$ is a mask variable defined by the indicator function $1[\cdot]$, reference emission timing $t_u$, left buffer $b_\texttt{l}$, and right buffer $b_\texttt{r}$ as
\begin{align}
    m_\phi(t, u) &\triangleq 1[t_u - b_{\texttt{l}} \leq t ~ \land ~ t < t_u + b_{\texttt{r}}],\\
    m_y(t, u) &\triangleq 1[t_u - b_\texttt{l} \leq t ~ \land ~ t \leq t_u + b_\texttt{r}].
\end{align}
The backward recursion in (\ref{eq:backward_recursion}) is also modified in the same manner. 

\subsection{FastEmit \cite{Yu21-FLL}}

FastEmit boosts the gradient of the label emission probability by a specified factor, so that those alignments that emit labels earlier are encouraged. Specifically, the gradient is increased by a factor of $(1 + \lambda_{\texttt{FE}})$ as
\begin{align}
\frac{\partial \mathcal{L}_{\texttt{FE}}}{\partial P_{y_{u+1}}} &\triangleq \big(1 + \lambda_{\texttt{FE}} \big) \frac{\partial \mathcal{L}_{\texttt{trans}}}{\partial P_{y_{u + 1}}}, \\
\frac{\partial \mathcal{L}_{\texttt{FE}}}{\partial P_{y_{\phi}}} &\triangleq \frac{\mathcal{L}_{\texttt{trans}}}{\partial P_{\phi}},
\end{align}
where $\lambda_{\texttt{FE}} \in \mathbb{R}_{>0}$ is the hyper-parameter to control the strength of the regularization. FastEmit does not require reference alignments, which makes the training pipeline much simpler. However, it cannot tell how much delay exists at each location $(t, u)$, and uniformly boosts the gradients at all locations regardless of the delays. This blind approach may make it difficult to achieve an optimal latency-accuracy trade-off.

\section{Minimum Latency Training}

\subsection{Expected latency} \label{sec:expected_latency}

\begin{figure}[t]
  \centering
  \includegraphics[width=8.5cm,angle=0]{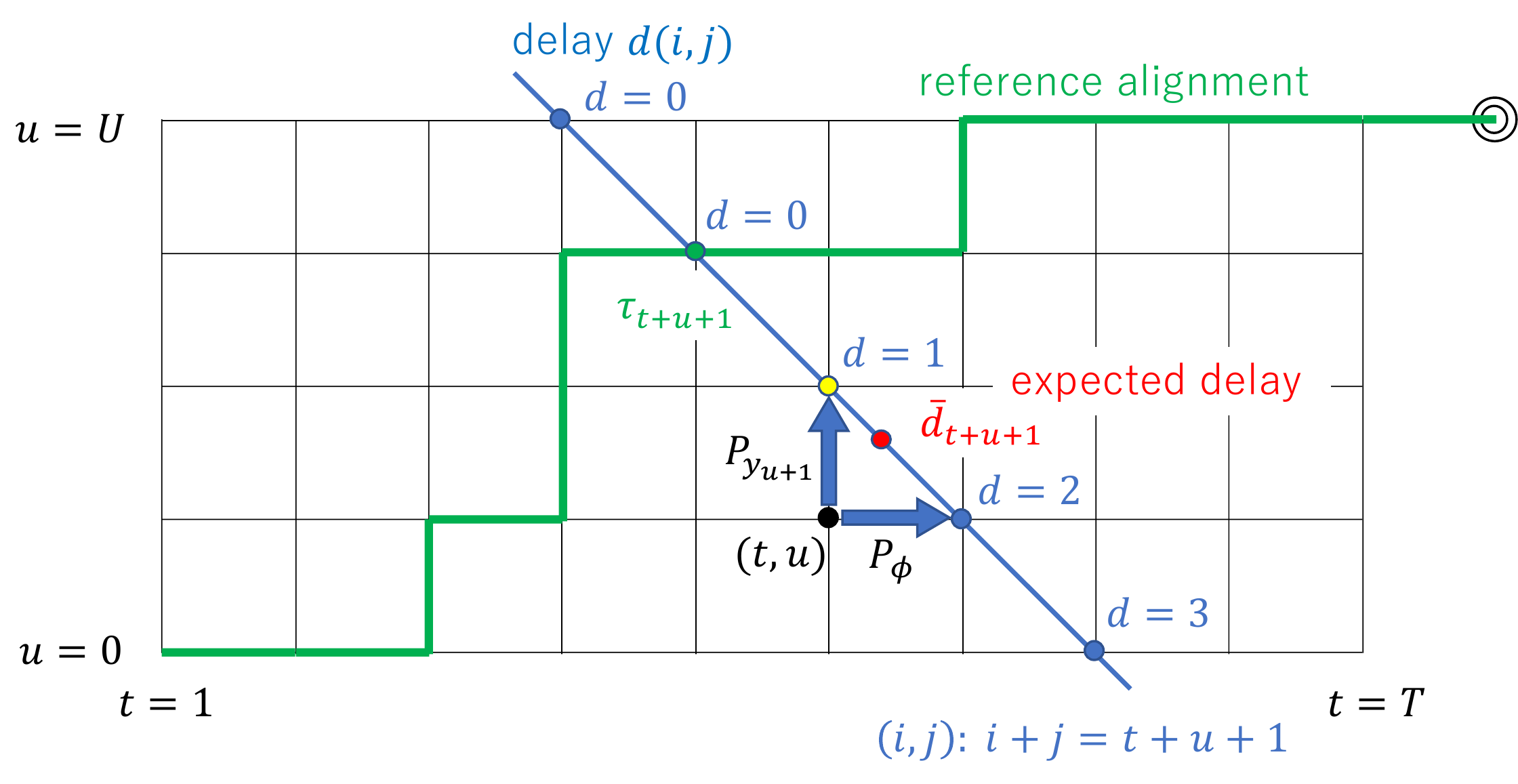}
  \caption{Delays $d(t, u)$ relative to the reference alignment are defined on the line $(t, u): t + u = n$ for $\forall n \in \{1, \dots, T+U+1\}$. Expected delay $\bar{d}_n$ can be calculated by averaging these delays weighted by the posterior probabilities as (\ref{eq:expected_delay}). The gradients of the expected delay with respect to the label and blank emission probabilities are calculated by (\ref{eq:grad_y}) and (\ref{eq:grad_b}).}
  \label{fig:delays}
\end{figure}

Let $d(t, u) \in \mathbb{R}_{\geq 0}$ be the delay at position $(t, u)$. Arbitrary values can be set as delays, but in this work we define them as,
\begin{equation}
d(t, u) \triangleq \max\Big(0, t - \tau_{\texttt{ref}}^{(t + u)}\Big).
\end{equation}
Here $\tau_{\texttt{ref}}^{(n)} \in \{1, \dots, T\}$ is the reference (ground-truth) time defined for $\forall n \in \{1, \dots, T+U+1\}$, and represents the reference alignment timing on the line $(t, u): t + u = n$. It can be generated for instance by forced-alignment using an HMM-hybrid model. Figure~\ref{fig:delays} depicts an example of a reference alignment and delays on the lattice.

Let $\bar{d}_n$ be an expected delay on the line $(t, u): t + u = n, \forall n \in \{1, \dots, T+U+1\}$ defined as
\begin{align}
\bar{d}_n &\triangleq \E_{(t, u): t + u = n} \Big[d(t, u)\Big] \\
  &= \sum_{(t, u): t + u = n} P(t, u | n, X) d(t, u) \\
  & = \sum_{(t, u): t + u = n} \frac{\alpha(t, u) \beta(t, u)}{P(Y | X)} d(t, u). \label{eq:expected_delay}
\end{align}
Here the posterior probability of being at position $(t, u)$ on line $(t, u): t + u = n$, denoted by $P(t, u | n, X)$ \footnote{It is similar to the notion of the state occupancy probability in the hidden Markov model.}, can be calculated using the forward and backward variables as described in (\ref{eq:expected_delay}).

The gradients of the expected latency $\bar{d}_{t+u+1}$ in (\ref{eq:expected_delay}) with respect to the label and blank emission probabilities, $P_{y_{u+1}} \triangleq P(y_{u+1} | t, u)$ and $P_{\phi} \triangleq P(\phi | t, u)$, can be calculated as \footnote{Derivation of the gradients is briefly explained in Appendix.}
\begin{align}
\frac{\partial \bar{d}_{t+u+1}}{\partial P_{y_{u + 1}}} &= \frac{\alpha(t, u) \beta(t, u + 1)}{P(Y|X)} \left( d(t,u + 1) - \bar{d}_{t + u + 1} \right), \label{eq:grad_y} \\
\frac{\partial \bar{d}_{t+u+1}}{\partial P_{\phi}} &= \frac{\alpha(t, u) \beta(t + 1, u)}{P(Y|X)} \left( d(t + 1,u) - \bar{d}_{t + u + 1} \right). \label{eq:grad_b}
\end{align}
An intuitive interpretation of the gradients in (\ref{eq:grad_y}) and (\ref{eq:grad_b}) is as follows. When the expected delay, represented by the red dot in Figure~\ref{fig:delays}, is located on the lower-right side of $(t, u + 1)$, represented by the yellow dot, $\bar{d}_{t+u+1}$ gets larger than $d(t, u+1)$ and the gradient in (\ref{eq:grad_y}) becomes negative; hence, label emission probability $P_{y_{u+1}}$ should be increased to reduce the expected delay, i.e.~to move the red dot to the upper-left direction. On the other hand, if the red dot is located on the upper-left side, the label emission probability should be decreased. The same argument can be made for the blank emission probability as well. So the gradients can be used to push the red dot towards the upper-left direction, i.e.~to reduce the expected latency of the sequence transducer models.

\subsection{Loss function}
The sequence transducer model tends to delay the alignment so that the model can access more future context to improve the loss function in (\ref{eq:transducer_loss}). In this work, we propose to augment the loss function with the expected delay introduced in Section~\ref{sec:expected_latency} so that the model learns to predict labels accurately and fast.
Specifically, the total loss is defined as the weighted sum of the transducer loss $\mathcal{L}_{\texttt{trans}}$ in (\ref{eq:transducer_loss}) and the expected latency $\bar{d}_{t+u+1}$ in (\ref{eq:expected_delay}) as
\begin{equation}
\mathcal{L}_{\texttt{MLT}} \triangleq \mathcal{L}_{\texttt{trans}} + \lambda_{\texttt{MLT}} ~ \bar{d}_{t+u+1},
\end{equation}
where $\lambda_{\texttt{MLT}} \in \mathbb{R}_{>0}$ is a hyper-parameter to control the strength of the regularization. By using (\ref{eq:trans_loss_grad_y}), (\ref{eq:trans_loss_grad_b}), (\ref{eq:grad_y}), and (\ref{eq:grad_b}), its gradients with respect to the label and blank emission probabilities can be calculated as
\begin{align}
\frac{\partial \mathcal{L}_{\texttt{MLT}}}{\partial P_{y_{u+1}}} = \frac{\partial \mathcal{L}_{\texttt{trans}}}{\partial P_{y_{u + 1}}} \bigg\{ 1 - \lambda_{\texttt{MLT}} \Big( d(t, u + 1) - \bar{d}_{t + u + 1} \Big) \bigg\}, \label{eq:mlt_loss_grad_y} \\
\frac{\partial \mathcal{L}_{\texttt{MLT}}}{\partial P_{y_{\phi}}} = \frac{\partial \mathcal{L}_{\texttt{trans}}}{\partial P_{\phi}} \bigg\{ 1 - \lambda_{\texttt{MLT}} \Big( d(t + 1, u) - \bar{d}_{t + u + 1} \Big) \bigg\}. \label{eq:mlt_loss_grad_b}
\end{align}
It can be seen from (\ref{eq:mlt_loss_grad_y}) and (\ref{eq:mlt_loss_grad_b}) that the gradient is discounted if the latency $d$ is larger than the expected latency $\bar{d}$, whereas the gradient is boosted if the latency is smaller than the expected latency, so that the alignments with lower latency are encouraged. It can also be seen that the blank gradient is more strongly discounted than the label gradient, since delay $d(t + 1, u)$ is larger than $d(t, u + 1)$ (typically by one), which means that label emission is encouraged in a similar manner as FastEmit. Note that FastEmit uniformly boosts the label gradient at all $(t, u)$ by a constant factor, while our proposed method changes the boosting (or discounting) factor depending on $(t, u)$ by considering the delays.

\section{Experiments}

Experiments were conducted on the Wall Street Journal dataset to compare minimum latency training with two conventional methods, namely alignment-restricted training and FastEmit, with respect to their accuracy and latency.

\subsection{Experimental settings}

The Wall Street Journal dataset \cite{Paul92-TDF} consisting of training set \texttt{si284} (37,416 utterances), development set \texttt{dev93} (503 utterances) and evaluation set \texttt{eval92} (333 utterances), which amount to 81 hours in total, was used to train and test the models.
Word Error Rate (WER) and Partial Recognition latency were used as the metrics of accuracy and latency, respectively. Specifically, Partial Recognition latency \cite{Yu21-FLL}, which represents the difference between the time when the user stops speaking and the time when the last token is emitted, at 50-th (PR50) and 90-th (PR90) percentiles of all utterances in the test set were used as the metrics of the latency.
Note that the times were measured in frame index $t \in \{1, \dots, T\}$, but reported in milliseconds by multiplying the frame period for better readability.

Causal Conformer transducer (Conformed-T) models consisting of a 12-layer/512-dim encoder, a 1-layer/512-dim LSTM-based prediction network, and a 512-dim joint network were used in this experiment. The encoder used Conformer layers with causal convolutions of kernel size 7, and 4-head self-attention modules that only attends to the past and present frames\footnote{The right context was masked while the full left context was used.}. Filter-bank features of 80-dim were extracted at a 10-ms frame shift, and fed into a 4-layer VGG-style input layers prepended to the encoder, where the features were downsampled by a factor of six to reach a frame period of 60 ms. SpecAugment was used to regularize the training. The model was configured to predict 52 tokens including the characters and the blank. The Adam optimizer was used in combination with the \textit{noam} learning-rate schedule, where 10,000 warmup steps was used. Inference was conducted using a standard beam search algorithm \cite{Graves12-STW} with a beam size of 10. No language model was used in this experiment. It is noted that the VGG-style input layers have a look-ahead of 120 ms in total, but this delay is not included in the reported PR50/PR90 numbers.

The causal Conformer-T models were trained with four different methods: the standard RNN-T training without any latency reduction technique (baseline) \cite{Graves12-STW}, alignment-restricted training \cite{Mahadeokar21-ARS}, FastEmit \cite{Yu21-FLL}, and the proposed minimum latency training. The reference alignments used for alignment-restricted training and minimum-latency training were generated by an HMM-hybrid model trained with Kaldi \cite{Kaldi}; it is noted that only word-level alignments were available, so the reference time of each character was set equal to the time of the word it belongs to. Regarding the alignment-restricted training, left buffer $b_\texttt{l}$ was set to 20, and right buffer $b_\texttt{r}$ was swept to draw a latency-WER tradeoff curve. As for FastEmit and minimum latency training, regularization strengths $\lambda_{\texttt{FE}}$ and $\lambda_{\texttt{MLT}}$ were swept, respectively. Other hyper-parameters (e.g.~learning-rate schedule and kernel size) were tuned on \texttt{dev93} for the baseline, and used by the other three methods without further tuning. Three independent training runs were conducted with different random seeds, and the average of the three trials are reported for each condition.
Experiments were conducted on ESPnet \cite{Watanabe18-EET,Guo21-RDE,Boyer21-ASO}.

\subsection{Results}
\begin{figure}[t]
  \centering
  \includegraphics[width=8.5cm,angle=0]{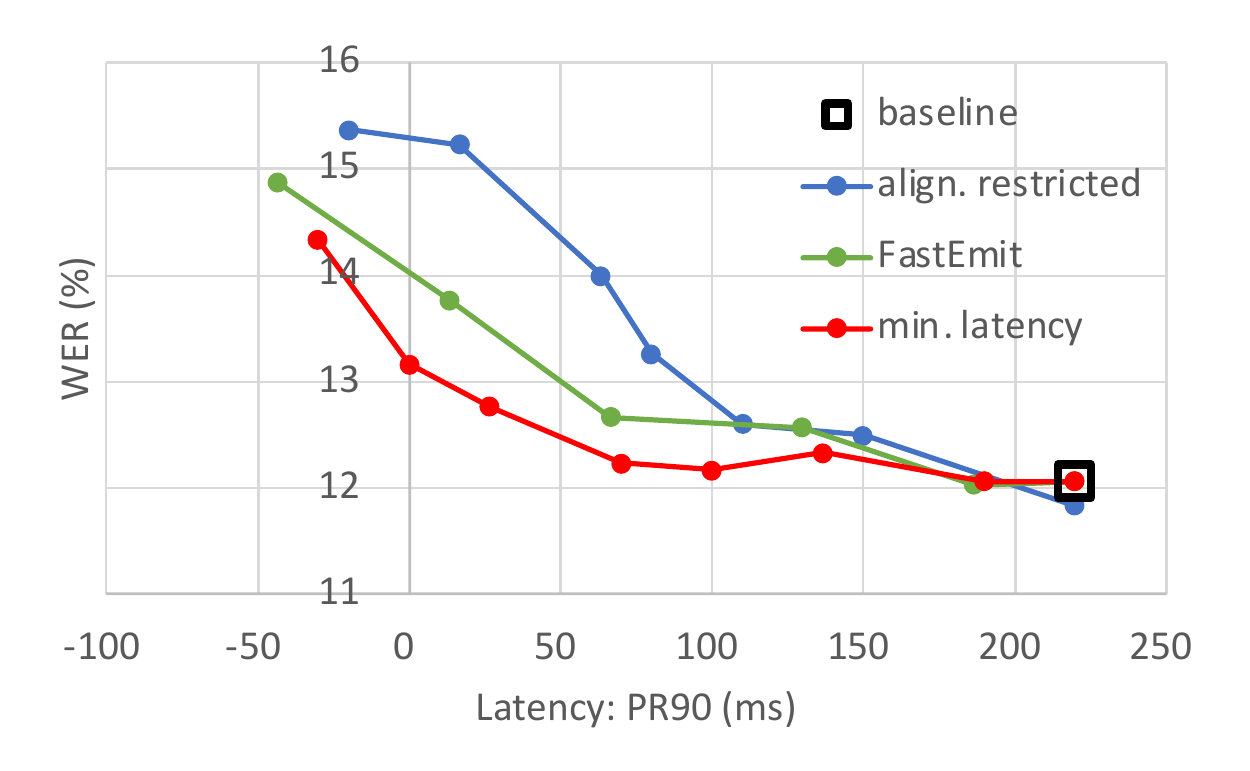}
  \vspace{-1em}
  \caption{WER (\%) and latency (ms) on \texttt{eval92} of WSJ with sweeping hyper-parameters $b_{\texttt{r}}$, $\lambda_{\texttt{FE}}$, and $\lambda_{\texttt{MLT}}$. Each point represents the average of three training runs. No language model was used.}
  \label{fig:wsj}
\end{figure}

Figure~\ref{fig:wsj} shows the results on WSJ. The baseline without using any latency reduction technique had a latency of 220 ms and a WER of 12.1\%. Although the model is causal and does not see any future context, the WER was close to that of the Transformer model (11.89\%) in \cite{Lohrenz21-RAS}, so the baseline should be reasonably strong.
Alignment restricted training successfully reduced the latency as $b_\texttt{r}$ was gradually reduced from 20 to 1, but the WER degraded significantly as a side effect.
FastEmit had a better latency-WER trade-off than alignment-restricted training, but still had a noticeable loss in WER especially when $\lambda_{\texttt{FE}}$ was large.
Minimum-latency training had the best latency-WER trade-off among the three methods, reducing the latency by more than 190 ms while holding the degradation in WER within 0.7\%. 

Table~\ref{tab:wsj} presents the results on WSJ with representative hyper-parameter settings to achieve a similar WER on $\texttt{dev93}$, namely $b_{\texttt{r}} = 9$, $\lambda_{\texttt{FE}} = 0.015$, and $\lambda_{\texttt{MLT}} = 0.03$. As was already seen in Figure~\ref{fig:wsj}, minimum latency training had the best latency-WER trade-off among the three methods in terms of PR90. However, the gap in PR50 between FastEmit and minimum latency training was relatively small compared with PR90. It is conjectured that the proposed method works to reduce the expected delay until it reaches zero, but does not push it any further by its design
(gradients in (\ref{eq:grad_y}) and (\ref{eq:grad_b}) approach zero), 
hence latency metrics do not take large negative numbers.

\begin{table}[t]
  \caption{WER (\%) and latency (ms) on WSJ with hyper-parameters chosen to achieve a similar WER on \texttt{dev93}. No language model was used.}
  \centering
  \begin{tabular}{l c c c c}
  \toprule
  \textbf{Method} & \textbf{WER} & \textbf{WER} & \textbf{PR50} & \textbf{PR90} \\
              & dev93\!\! & eval92\!\! & eval92\!\! & eval92\!\! \\
  \midrule
  Conformer-T (\texttt{Base}) & 14.6 & 12.1 & 143 & 220 \\
  \midrule
  \texttt{Base} + align.~restr'd & 15.9 & \textbf{12.6} & 33 & 110 \\
  \texttt{Base} + FastEmit & 16.0 & 12.7 & -50 & 67 \\
  \texttt{Base} + min.~latency & \textbf{15.7} & 12.8 & \textbf{-53} & \textbf{27} \\
  \bottomrule
  \end{tabular}
  \label{tab:wsj}
\end{table}

\section{Conclusions}

In this paper, we proposed minimum latency training of sequence transducer models for fast and accurate end-to-end speech recognition. We have defined the expected latency on each diagonal line on the lattice, and shown that its gradients can be computed efficiently within the forward-backward algorithm.
In experiments on the WSJ dataset, the proposed minimum latency training consistently had a better latency-accuracy trade-off over a wide range of hyper-parameters compared with alignment-restricted training and FastEmit. With a representative hyper-parameter, the proposed method reduced the latency of a causal Conformer-T model from 220 ms to 27 ms within a WER degradation of 0.7\%, and outperformed alignment-restricted training (110 ms) and FastEmit (67 ms).

\section{Appendix}

In this section, we briefly explain the derivation of the gradient of $\bar{d}_{t+u+1}$ with respect to the label emission probability $P_{y_{u+1}} \triangleq P(y_{u+1}|t,u)$ in (\ref{eq:grad_y}). The gradient w.r.t.~the blank emission probability in (\ref{eq:grad_b}) can be also derived in a similar manner.
By recalling that for $\forall (i,j): i+j=t+u+1$,
\begin{align}
    \frac{\partial \alpha(i,j)}{\partial P_{y_{u+1}}} &= 
    \begin{cases}
      \alpha(t, u),&\text{if } (i,j) = (t, u+1)\\
      0,&\text{otherwise}
    \end{cases} \\
    \frac{\partial \beta(i,j)}{\partial P_{y_{u+1}}} &= 0,
\end{align}
and that the following equality holds for $\forall n \in \{1, \dots, T+U+1\}$,
\begin{equation}
    P(Y|X) = \sum_{(i,j): i+j=n} \alpha(i,j) \beta(i,j),
\end{equation}
the gradient can be calculated as
\begin{align}
    \frac{\partial \bar{d}_{t+u+1}}{\partial P_{y_{u+1}}} &= \frac{\alpha(t,u)\beta(t,u+1)}{P(Y|X)^2} \Big\{ P(Y|X)d(t,u+1) \nonumber \\
    &~~~~~ - \sum_{(i,j):i+j=t+u+1}\alpha(i,j) \beta(i,j) d(i,j) \Big\} \\
    &= \frac{\alpha(t,u)\beta(t,u+1)}{P(Y|X)} \Big\{ d(t,u+1) - \bar{d}_{t+u+1}\Big\}.
\end{align}

\bibliographystyle{IEEEtran}
\bibliography{mybib}

\end{document}